# Five Clarifications about Cultural Evolution

Liane Gabora

For correspondence regarding the manuscript:
Liane Gabora
Department of Psychology, University of British Columbia
Okanagan Campus, 3333 University Way
Kelowna BC, Canada V1V 1V7
Email: liane.gabora@ubc.ca
Tel: 250-807-9849
Fax: 250-470-6001

ABSTRACT

This paper reviews and clarifies five misunderstandings about cultural evolution identified by Henrich, Boyd, and Richerson (2008). First, cultural representations are neither discrete nor continuous; they are distributed across neurons that respond to microfeatures. This enables associations to be made, and cultural change to be generated. Second, 'replicator dynamics' do not ensure natural selection. The replicator notion does not capture the distinction between actively interpreted self-assembly code and passively copied self-description, which leads to a fundamental principle of natural selection: inherited information is transmitted, whereas acquired information is not. Third, this principle is violated in culture by the ubiquity of acquired change. Moreover, biased transmission is less important to culture than the creative processes by which novelty is generated. Fourth, there is no objective basis for determining cultural fitness. Fifth, the necessity of randomness is discussed. It is concluded that natural selection is inappropriate as an explanatory framework for culture.

It is indisputable that elements of culture adapt over time, and exhibit phenomena observed in biological evolution such as niches, drift[1], epistasis[2], and punctuated equilibrium[3] (Bentley et al., 2004; Durham, 1991; Gabora, 1995, Mesoudi, Whiten & Laland, 2004, 2006; Orsucci, 2008). It has been argued that culture constitutes a second evolutionary process, one which, though it grew out of biological evolution, exhibits an evolutionary dynamic in its own right that cannot be reduced to biology. Care must be taken because at least some of what is considered cultural behavior *can* be accounted for by a purely biological explanation (Tooby & Cosmides, 1992). However most would probably concede that, much as principles of physics do not go far toward an explanation of, say, the giraffe's long neck (though things like gravity play some role), biology does not go far toward an explanation of, say, the form and content of a poem (though factors like selective pressure for intelligence play some role). Most would concede that to explain how and why such forms arise, accumulate, and adapt over time, one must look to culture. Nevertheless, the attempt to establish a convincing evolutionary framework for culture remains a struggle.

Symptomatic of this struggle is a paper that addressed five misunderstandings concerning efforts to frame culture in evolutionary terms (Henrich, Boyd, & Richerson, 2008). The present paper builds on this initiative, addressing these same five issues, and clarifying further sources of misunderstanding arising in the original paper. Specifically, it discusses the nature of mental representations and biases in how they are transmitted, whether natural selection entails replicators, the cultural equivalent of fitness, and whether variation need be random for natural selection to be applicable. The paper concludes by sketching a theory that culture evolves through processes similar to those by which the earliest life forms evolved prior to emergence of the genetic code.

## Mental Representations are Neither Discrete Nor Continuous but Distributed

The first misunderstanding concerns the nature of mental representations. Some claim that models of cultural evolution in which mental representations assume a discrete, gene-like, particulate structure are useless (Atran, 2001), while Henrich et al. (2008) point to results obtained with one such model as evidence that they are valuable.

Let us look carefully at both sides. The fact that genes are particulate enables genetic diversity to be maintained. If they were not particulate, then variation would become diluted with time (much as mixing black and white paint gives gray paint; there is no way to 'get back' the pure black or white paint.) Memeticists argue that cultural representations, or memes, must be particulate, as are genes (Blackmore, 1999; Aunger, 2000). Many models of cultural evolution *do* assume particulate representations (e.g. Boyd & Richerson, 1985; Cavalli-Sforza & Feldman, 1981; Henrich & Boyd, 1998). However, mental representations are *not* particulate (Goldman-Rakie, 1992; Miyake & Shah, 1999).

---

[1] Drift refers to changes in the relative frequencies of variants through random sampling from a finite population. It is the reason why variation is reduced in small, reproductively isolated populations. Drift has been shown to occur in a cultural context with respect to such things as baby names and dog breed preferences (Neiman 1995; Madsen et al. 1999; Bentley et al. 2004). Drift is also observed in EVOC, a computer model of cultural evolution (Gabora, 1995, 2008a,b).

[2] In biology epistasis refers to the situation where the fitness at one gene locus depends on which allele (version of a gene) is present at another locus. When applied to culture it refers to the situation where the value of a particular idea or act depends on what other ideas one holds, or acts one engages in (Gabora, 1995, 2008a,b).

[3] Punctuated equilibrium is the theory that sexually reproducing populations generally experience little change for most of their history, but when change does occur it is localized in rare, rapid 'branching' events (Eldredge & Gould, 1972; Gould & Eldredge, 1993). There is evidence for an analogous pattern in culture (Orsucci, 2007).

It has been argued that they are instead graded or continuous, and that because of this, cultural traits—unlike biological traits—can blend (Atran, 2001).[4] There exist mathematical models of culture that achieve blending by allowing mental representations to take on continuous values. Henrich et al. (2008) point to one such model (Henrich & Boyd, 2002) in which each agent's belief is represented as a numerical value ($x$) between zero and one. Belief $x = 0$ might be that the moon has emotions and its color expresses its mood, while $x = 1$ represents the belief that the moon is simply a big rock, without emotions, and its color is due to the laws of light refraction. They give as an example that is intermediate between these extremes the belief that the moon's color is 23% controlled by its emotions and 77% controlled by the laws of refraction, noting "such beliefs might seem odd to us because they violate intuitive expectations, which is why cognitive attractors might transform them" (p. 122). The beliefs $x = 0$ and $x = 1$ are referred to as *cognitive attractors* (following Sperber's (1996) use of the term) because they are easier to think. In the model, the agents' beliefs do indeed shift over time from intermediate values to either 1 or 0. Unfortunately, beliefs of the sort 'the moon's color is 23% controlled by its emotions and 77% controlled by the laws of refraction' are not just odd, they are psychologically unrealistic, and moreover the way people's representations generally change over time is not from complicated to simple but in the other direction, from simple to more complex and nuanced reflections of their world (see also, Claidière & Sperber, 2007). Thus the model is unconvincing as an example of gaining a better understanding of how culture evolves by paying close attention to the psychological processes involved.

There *is* a sense in which the discrete view of mental representations is correct. A neuron either fires or does not, and at this basic level mental representation involves discrete processes. However, it is not because mental representations are continuous that they are subject to change, but because they are *distributed* across cell assemblies of neurons. Knowledge is encoded in neurons that are sensitive to high level features or low level *microfeatures* (Churchland & Sejnowski, 1992; Smolensky, 1988). For example, one might respond to a particular shade of red, or a salty taste, or something that does not exactly match an established term (Mikkulainen, 1997). Although each neuron responds maximally to a particular microfeature, it responds to a lesser extent to related microfeatures, an organization referred to as *coarse coding*. For example, neuron *A* may respond preferentially to lines of a certain angle from the horizontal (say 90 degrees), while its neighbor *B* responds preferentially to lines of a slightly different angle (say 91 degrees), and so forth. However, although *A* responds *maximally* to lines of 90 degrees, it responds somewhat to lines of 91 degrees. Thus the encoding of a mental representation is distributed across one or more cell assemblies containing many neurons, and likewise, each neuron participates in the encoding of many items (Hinton, McClelland, & Rumelhart, 1986). The same neurons get used and re-used in different capacities, a phenomenon referred to as *neural re-entrance* (Edelman, 1987). Items stored in overlapping regions are correlated, or share features. Memory is also *content addressable* in that there is a systematic relationship between the content of an input and which particular neurons encode it[5]. As a result, items in memory can thereafter be evoked by stimuli that are similar or 'resonant' (Hebb, 1949; Marr, 1969).

---

[4] Actually, blending is possible with discrete traits. Indeed it is common in biology for polygenic traits (traits coded for by more than one gene), the classic example being Mendel's cross of red and white flowers to yield pink ones.

If the regions in memory where two distributed representations are encoded overlap, then they share one or more microfeatures. They may have been encoded at different times, under different circumstances, and the relationship between them never explicitly noticed. But the fact that their distributions overlap means it is possible that some situation could occur that would cause one to evoke the other, enabling one to be seen from the perspective of the other, thereby potentially modifying it. There are as many ways of generating associations and reminding events as there are microfeatures by which they overlap; *i.e.* there is room for atypical as well as typical associations. It is because the region of activated neurons is distributed, but not *too* widely distributed, that one can generate a stream of coherent yet potentially creative thought. The more detail with which items have been encoded in memory, the greater their potential overlap with other items, and the more routes to forging relationships between what is currently experienced and what has been experienced before.

The upshot is that *it is not because representations are continuous that beliefs and ideas change over time, it is because they are distributed and content-addressable.* Like particulate models, models based on distributed representations can retain diversity, but like continuous models, they can capture gradations, and moreover they can begin to capture how ideas change when we think them through, combine them, and discuss them. There exist models of cultural evolution in which mental representations are neither discrete nor continuous but distributed. For example, in EVOC, a computer model of cultural evolution composed of a society of artificial agents without birth, death, or genomes, the agents' actions improve over time through the invention of new actions and imitation of neighbors' actions. Ideas for actions are distributed across mental representations of their body parts, and invention makes use of hunches that build up over time about what makes for effective actions (Gabora, 1995, 2008a,b).

---

[5] Most computers are content addressable, but they work differently. Each different input is stored in a unique address. Retrieval is a matter of looking at the address in the address register and fetching the item at the specified location. Since there is no *overlap* of representations, there is no means of forging associations based on newly perceived similarities. The exception is computer architectures that exhibit sparseness, distributed representation, and content addressability (Kanerva, 1988). A connectionist memory is able to abstract a prototype, fill in missing features of a noisy or incomplete pattern, or create a new pattern on the fly that is more appropriate to the situation than anything it has ever been fed as input (Rummelhart & McClelland, 1986). Like a human memory it does not retrieve an item so much as reconstruct it.

## Replicators, Replicator Dynamics, and Self-replicating Automata

The second misunderstanding about cultural evolution addressed by Henrich et al. is the claim that replicators are necessary for Darwinian evolution (Dawkins, 1976, 1982). A replicator, according to Dawkins, is "any entity in the universe which interacts with its world, including other replicators, in such a way that copies of itself are made" (Dawkins, 1976; p. 17). A replicator is said to have the following properties:

- *Longevity*—it survives long enough to replicate, or make copies of itself.
- *Fecundity*—at least one version of it *can* replicate.
- *Fidelity*—even after several generations of replication, it is still almost identical to the original.
-

Dawkins suggested that replicators can be found not just in biology but also in culture, and he christened these cultural replicators 'memes'. He elaborates: "Just as genes propagate themselves in the gene pool by leaping from body to body via sperm or eggs, so memes propagate themselves in the meme pool by leaping from brain to brain."

The replicator notion was not unequivocally welcomed by biologists. Ernst Mayr, one of the foremost evolutionary biologists of the 20th century, claimed that the replicator notion is "in complete conflict with the basics of Darwinian thought" (1996: p. 2093). According to Stuart Kauffman, the replicator concept is impoverished because it does not capture the essential features of the kind of structure that evolves through natural selection (Kauffman, 2009). However, the replicator notion was embraced by some cultural theorists, who came to equate the claim that culture evolves through natural selection with commitment to the notion of a 'second replicator' (Aunger, 2000; Blackmore, 1999; Dennett, 1995). The assumption is that if a 'cultural replicator' is found we will have proof that culture evolves through natural selection. Henrich et al. (2008) disagree, pointing to evidence that replicators are *sufficient* for natural selection, but not *necessary* (Henrich & Boyd, 2002). Let us examine this claim.

## The Abstract Structure of a Natural Selection Process

To assess claims about replicators it is necessary to look at the basic algorithmic structure of natural selection. Perhaps no one has a better understanding of this than John Holland, inventor of the genetic algorithm (Holland, 1975). A genetic algorithm is a computer program that solves complex problems by randomly generating variant solutions and selectively 'mating' the best for multiple 'generations' until an adequate solution is found. Holland, who was profoundly influenced by the formal analysis of the abstract structure of living things in John von Neumann's (1966) book 'Theory Of Self-Reproducing Automata', cites three fundamental, interrelated principles of natural selection: (1) sequestration of inherited information, (2) clear-cut distinction between genotype and phenotype, and (3) no transmission of acquired traits. Chris Langton, who initiated the field of artificial life, points to the same three principles (Langton, 1992). By examining them we can better ascertain the extent to which the replicator notion captures the essence of natural selection.

The concept of sequestration of inherited information is most easily appreciated in relation to the paradox it resolved. The paradox faced by Darwin and his contemporaries was the following: how does biological change *accumulate* when traits acquired over an organism's lifetime are *obliterated?* For example, a rat whose tail is cut off does not give birth to rats with cut-off tails; the rat lineage loses this trait. Darwin's genius was to explain how living things adapt over time despite that new modifications keep getting discarded, by looking from the level of the individual to the level of the *population* of interbreeding individuals. He realized that individuals who are better equipped to survive in their given environment leave more offspring (are 'selected'). Thus, although their *acquired* traits are discarded, their *inherited* traits (often the traits they were born with) are more abundant in the next generation. Over generations this can lead to substantial change in the distribution of traits across the population as a whole.

Although Darwin observed that this was the case, he did not know why. We now know that the reason acquired traits are discarded is that organisms replicate not by mere self-copying, but self-copying using a coded set of self-assembly instructions that gets used in two distinct ways. One way it is used involves actively deciphering the coded instructions to construct a highly self-similar 'variant'. In this case, the code functions as *interpreted* information. The second way is as a self-description that is passively copied to the replicant to ensure that *it* can reproduce. In this case, the self-assembly instructions function as *un-interpreted* information. (To put it more loosely, the interpreting can be thought of as 'now we make a body', and the un-interpreted use of the code as 'now we make something that can *itself* make a body'.) Von Neumann refers to a structure that evolves through natural selection as a *self-replicating automaton*.

Since biology is the field that inspired this distinction, naturally it applies here. The DNA self-assembly code is copied—without interpretation—to produce gametes during meiosis. If gametes unite to form a germ cell, their DNA is decoded—interpreted—to synthesize the proteins necessary to assemble a body. The process by which this unfolds is referred to as development. The DNA self-assembly code, i.e. the information that gets inherited, is said to be *sequestered* because it is largely shielded from environmental influence during its lifetime.

The distinction between these two ways of using a self-assembly code, one enabling gamete production and the other enabling development, leads to another distinction: the distinction between genotype and phenotype. The *genotype* of an organism is its genetic constitution: the set of genes it inherited from its parents. The *phenotype* of an organism is its observable characteristics or traits, including morphological, developmental, biochemical, physiological, and behavioral traits. Phenotypes result through the expression of an organism's inherited genotype in the context of its environment. Phenotypes are altered—they acquire traits—through interaction with an environment, while genotypes generally do not.

Thus natural selection incorporates not just a means by which *inherited* variation is *passed on* (*e.g.,* you may have inherited your mother's blue eyes), but also a means by which variation *acquired* over a lifetime is *discarded* (*e.g.,* you did not inherit your mother's tattoo). What gets transmitted from parent to offspring is sets of self-assembly instructions, and these instruction sets have not changed since they were formed. So traits acquired by parents during their lives, including learned behavior and knowledge, are not inherited by offspring. Since acquired change can accumulate orders of magnitude faster than inherited change, if it is *not* getting regularly discarded, it quickly overwhelms the population-level mechanism of change identified by Darwin; it 'swamps the phylogenetic signal'. Thus natural selection only works as an explanation when acquired change is negligible.

Indeed what *necessitated* the theory of natural selection, a theory of *population-level* change, is that acquired traits are (with few exceptions) not inherited from parent to offspring. Note how exceptional this is. In most domains of human inquiry, change is retained, and successive changes accumulate. If an asteroid crashes into a planet, the planet cannot revert to the state of having not had the asteroid crash into it. But in the biological domain where, if a rat loses its tail the rat's offspring are *not* born tail-less, how does one explain how change is retained and accumulates? That was the extraordinary paradox Darwin faced, the paradox for which natural selection provided a solution. An analogous paradox does not exist with respect to culture.

**Do Replicators Embody the Abstract Structure of Natural Selection?**

Having examined the algorithmic structure of natural selection let us see how well it is embodied by cultural evolution. Even speaking loosely or metaphorically, an idea or artifact or meme cannot be said to possess a code that functions as both a passively copied self-description and a set of actively interpreted self-assembly instructions (Gabora, 2004). Not only have we no neurological evidence of this, but daily life provides a continuous stream of counter-evidence. We saw that the sequestered self-assembly code is vital to natural selection, and that the telltale signature of replication by way of a self-assembly code is lack of transmission of acquired characteristics. Therefore, if culture *did* evolve through natural selection, if you heard an idea from a friend, and you reframed it and modified it and put it in your own terms before expressing it to me, none of this reframing and modification—i.e. in biological terms, none of these acquired traits—would be present in the version you expressed to me. If culture evolved through natural selection, I would get it in exactly the form your friend expressed it to you. But this is clearly not what happens. Transmission of acquired traits is ubiquitous in culture. This paper is an example. Its contents got mulled over, literature was incorporated… information was acquired and transmitted. In culture there *is* no distinction between inherited and acquired traits; *all* traits are acquired. The use of the term 'dual inheritance' to refer to both what is transmitted genetically and what is transmitted culturally is technically incorrect and misleading. That which is transmitted through culture falls under the category of acquired change, not inherited change.

Lake (1998) argues that some but not all socially transmitted ideas and artifacts are replicators. The argument begins with a distinction between the *expression* versus the symbolically coded *representation* of cultural information. Whereas, for example, singing a song is an expression of a musical concept, a musical score is a representation of it. As another example, the spontaneous verbal explanation of an idea is an expression, whereas the text version of it is a representation. Lake comments that some cultural entities, such as village plans, constitute *both* a representation of a symbolic plan, and an expression of that plan, because they are both expressed by and transmitted through the same material form. However neither the expression nor the representation of a plan involves either interpreting or uninterpreted copying of a *self-assembly code*. A village plan does not, on its own, produce little copies of itself. A musical score does not generate 'offspring scores'. The perpetuation of structure and the presence of a symbolic code do not guarantee the presence of a self-replicating automaton. Symbolic coding is not enough; it must be a coded representation of the *self*.

This distinction was recognized by Maturana and Varela (1980), who use the term *allopoietic* to describe an entity (such as a village plan) that generates another entity (such as a village) with an organization that is different from its own (e.g. while a village plan is a two-dimensional piece of paper, it generates something that is three dimensional and constructed from a variety of materials). Maturana and Varela contrast this with *autopoietic*, a term used to describe an entity composed of parts that through their interactions regenerate themselves and thereby reconstitute the whole. It thereby forges another entity with an organization that is nearly identical to its own. A self-replicating automaton is autopoietic. Socially transmitted elements of culture are not.

Let us now return to the debate about replicators. Henrich et al claim that replicators are sufficient but not necessary for natural selection on the basis of a mathematical model of culture that exhibits cumulative change without any self-replicating structure (Henrich & Boyd, 2002). Equating cumulative change with 'replicator dynamics', and 'replicator dynamics' with natural selection, they further claim that self-replication is not necessary for natural selection. Therefore, they conclude, any lack of self-replication in culture does not constitute a valid argument against the view that culture evolves through natural selection. Where they go wrong is in assuming that 'replicator dynamics' is a litmus test for natural selection. The set of models that can exhibit replicator dynamics includes models, such as Henrich and Boyd's, that lack the key features of natural selection: sequestering of inherited information, genotype/phenotype distinction, and no transmission of acquired traits. Their claims are made on the basis of a model of cultural evolution in which the algorithmic structure of natural selection is reduced to mere cumulative change.

*The bottom line is that in order to understand in what sense culture evolves we must look beyond replicators and 'replicator dynamics'.* Contrary to Henrich and Boyd's claim, replicators are *necessary* for natural selection but not *sufficient*. The replicator concept is fine so far as it goes, but it does not capture the algorithmic structure of replication through natural selection. The fundamental structural feature of entities that evolve through natural selection is the distinction between actively interpreted self-assembly code and passively copied self-description, which leads to the following dynamical feature: inherited information is transmitted whereas acquired information is not. Because this is not the case with respect to culture, culture does not display the evolutionary dynamics of a Darwinian process, and Darwinian models of culture stand on weak theoretical grounds. Epidemiological treatments of culture are similarly flawed (their 'infectiousness' notwithstanding) because since viruses replicate by way of a self-replication code (although not exclusively their own), they too are by and large shielded from inheritance of acquired traits.

### Darwinian Approaches are Forced to Ignore Creativity and Contemplation

The third misunderstanding pointed to by Henrich et al. is Sperber's (1996) assertion that content-dependent psychological biases are the only important processes that affect the spread of cultural representations. By "content-dependent psychological biases" they mean the creative, strategic, intuitive processes that affect the actual content of ideas. The psychological processes that cultural Darwinists emphasize are those that affect not the content of ideas but the probability that they are imitated. These processes include *conformity bias,* which occurs when people preferentially adopt widespread cultural variants over rare ones, *prestige bias,* which occurs when people preferentially imitate high status individuals, and *copying error* (Boyd & Richerson, 1985; Richerson & Boyd, 2005). Darwinian approaches to culture avoid incorporating any sort of creative or contemplative thought processes[6]; ideas that exist in brains as a result of thinking something through for oneself have no place in their models. Their emphasis is squarely on biases in social transmission: "[culture consists of] information stored in brains—information that got into those brains by various mechanisms of social learning…" (Henrich et al., 2008: p. 120).

Henrich et al. overstate Sperber's position (see Claidière & Sperber, 2007); I suspect few would argue that transmission biases play no role at all in culture. There are subtler yet more fundamental problems with the cultural Darwinist's focus on transmission bias: (1) transmission biases, while important, are less important to culture than the strategic, creative processes that generate and modify cultural content in the first place, and (2) these creative processes *cannot* be accommodated by a Darwinian approach to culture because they entail retention of acquired characteristics. Since acquired change drowns out the slower inter-generational mechanism of change identified by Darwin (as explained earlier), natural selection is only of explanatory value when there is negligible transmission of acquired characteristics. This *is* the case in biological evolution, as we saw with the cut-off tail example; change acquired during an individual's lifetime is not generally passed on. As another example, you did not inherit your mother's knowledge of politics, or her tattoo, both of which she acquired between the time she was born and the time she transmitted genetic material to you. However, no serious scholar is likely to accept that there is negligible transmission of acquired characteristics in culture.

The only way to maintain a Darwinian perspective on culture that is consistent with the algorithmic structure of natural selection is to view humans as passive imitators and transmitters of prepackaged units of culture, which evolve as separate lineages. To the extent that these lineages 'contaminate one another'; that is, to the extent that we actively transform elements of culture in ways that reflect our own internal models of the world—altering or combining them to suit our needs, perspectives, or aesthetic sensibilities—natural selection cannot explain cultural change. In order for something to stick in memory we first relate it to what else we know, make it our own (Piaget, 1926), but a Darwinian framework for culture is incompatible with this (Sperber, 1996). It cannot begin to account for what happens when artists take familiar objects and themes and rework or juxtapose them, forcing us to reconceptualize them, to adopt a richer appreciation of their meaning in different contexts. Humans exhibit to a surprising degree the tendency to go our own way and do our own thing. Our noteworthy cultural achievements—the Mona Lisa, the steam engine, quantum mechanics—are widely known about and appreciated, but few attempt to imitate them, and moreover it is not those who merely imitate them that make cultural history but those who deviate from what has come before. As one choreographer (whose name I forget) put it: "If we don't do what our predecessors did, we're doing what our predecessors did." By placing undue emphasis on social learning, Darwinian models ignore exactly those factors that exert the greatest impact on culture. Indeed, creative individuals are the least socially tethered of all, with strong leanings toward isolation, nonconformity, rebelliousness, and unconventionality (Crutchfield, 1962; Griffin & McDermott, 1998; Sulloway, 1996). It has been proposed that what is unique to the human species, and what enabled human culture to become an *evolutionary* process, was onset of not the ability to imitate or influence others but the capacity to take an idea and reframe it in our own terms, to put our own spin on it (Gabora, 2003, 2008).

---

[6] Some have attempted to apply Darwinian thinking to the creative thought processes occurring within an individual, referring to the creative process as a process of Blind Variation and Selective Retention, or BVSR (Campbell 1960; Simonton 1998, 1999a,b, 2007a,b). The attempt to account for creativity in Darwinian terms has not been accepted by the scientific community primarily on the grounds that it is theoretically unsound, but for other reasons as well (Dasgupta, 2004; Eysenck, 1995; Gabora, 2005, 2007; Sternberg, 1998, Thagard, 1980; Weisberg, 2000, 2004, Weisberg & Haas, 2007). The problems with applying Darwinism to culture discussed in the present paper apply also to BVSR. Natural selection was not put forward to explain how biological novelty *originates*. It *assumes* random variation of heritable traits, and provides an explanation for population-level change over generations in the *distribution* of variants. I will not elaborate on this further here since it is not central to the current paper but point the reader to the above-mentioned papers where it is dealt with at length.

The fact that memory is distributed and content-addressable is critically important for the creative processes by which mental representations acquire change (Gabora, 2002, 2010). In a sparse, distributed, content-addressable memory, associations between items are made, not by chance, but because the items share (high-level or low-level) features. Even if the two items were encoded at different times and the relationship between them never explicitly noted before, because they are encoded in overlapping distributions of neurons, in a new context the connection between them may suddenly come to light. If it were not distributed, there would be no overlap between items that share features, and thus no means of forging associations amongst them. If it were not content-addressable, associations would not be meaningful. Content addressability is why the entire memory does not have to be searched or randomly sampled; it ensures that one naturally retrieves items that are relevant to the current goal or experience, sometimes in a rarely noticed but useful or appealing way. Actually, the notion of retrieving from memory is somewhat inaccurate; one does not retrieve an item from memory so much as *reconstruct* it (Edelman, 2000). This is why an item in memory is never re-experienced in exactly the form it was first experienced, but colored, however subtly, by what has been experienced in the meantime, and spontaneously re-assembled in a way that is relevant to the task at hand.

In sum, *the reason that Darwinian theories of culture do not incorporate phenomena such as creativity is not just because these phenomena are difficult and they have not got around to it yet*. *It is because a Darwinian framework cannot encompass them*. Cultural Darwinists limit the psychological processes in their models to factors that bias 'who imitates who' because if one faithfully imitates an idea, one does not put one's own slant on it—it does not acquire characteristics—and thus a Darwinian explanation is not rendered invalid.

## Cultural Fitness and Hitckhiking

The fourth claim made by Henrich et al. (2008) in 'five misunderstandings about cultural evolution' is that the cultural fitness of a mental representation cannot be inferred from its successful transmission through a population. This is true.[7] Henrich et al. do not cite any actual instances of cultural fitness being inferred from its successful transmission, and thus the issue may be something of a straw man, but it is worthy of discussion. Henrich et al. correctly point out that deleterious ideas or practices may spread because they are correlated with attractive individuals or successful groups. They may also simply be correlated with beneficial ideas or practices. For example, we engage in the practice of eating Chinese food, which is beneficial, except that it sometimes has MSG in it, which is deleterious. The practice of consuming MSG, though deleterious, spreads because it is correlated with the practice of eating Chinese food, which is beneficial. In biology the term for this is *hitchhiking* (Kaplan, Hudson, & Langley, 1989; Kojima & Schaeffer, 1967; Maynard Smith & Haigh, 1974). The closer together genes are on a chromosome the less likely they will be separated by crossover, so the more tightly linked they are. Hitchhiker alleles confer no fitness advantage, but endure because they are linked to alleles that are important for survival. Hitchhiking also occurs in culture (Gabora, 1996). For example, we can say that the practice of consuming MSG is hitchhiking on the practice of eating Chinese food. In both genetic hitchhiking and its cultural analog, useless (or even detrimental) patterns proliferate through their association with beneficial ones.

Although it is correct that the cultural fitness of a mental representation cannot be inferred from its successful transmission, a more pressing issue is whether the cultural fitness of a mental representation can be determined at all. Let us examine two serious problems in applying the concept of fitness to culture.

### Lack of 'Generations'

In biological terms, the fitness of an organism is the number of offspring it has in the next generation. The term 'generation' is applicable when individuals are irretrievably lost from a population and replaced by new ones. But with respect to culture, an outdated (or seemingly 'dead') idea or artifact can come back into use (or seemingly 'come back to life') when styles change or circumstances become right. Because there is no hard and fast distinction between a living entity and a dead one, there is no basis for determining what constitutes a generation. Thus the term generation does not apply to culture, and this is also the case with respect to the earliest stage of biological life (Gabora, 2006; Vetsigian et al, 2006).

### Ambiguous Ancestry

Second, Darwinian approaches assume that it is possible to accurately and objectively measure the relatedness of elements of culture. Whether or not two organisms share a common ancestor is clear-cut; they either are or are not descendents of a particular individual. One can objectively measure what percentage of the genomes of two species or two organisms overlap, and make conclusions about whether one is descended from another, and their degree of genetic relatedness. In a cultural context, however, there is no objective way to go about this. For example, consider the following unusual artifact: a couch with a ledge-shaped seat and letter-shaped pillows that was inspired by the game of Scrabble (fig. 1). Is this furniture descended from the game of Scrabble? Or was the Scrabble ledge which 'seats' a player's letters originally descended from the concept of a couch or chair? Did Scrabble somehow introduce a 'mutated taxon' into the 'furniture' lineage, or had the concept of an object to sit upon already contaminated the 'game' lineage when the inventor of Scrabble came up with the idea of perching letters on a ledge? What exactly was inherited?

---

[7] Thus their paper is mis-named; they only display four misunderstandings, not five.

[INSERT FIGURE 1 ABOUT HERE]

Henrich et al. speak of the differential fitness of cultural variants, but on what basis is one idea a variant of another? In a world where thoughts of chairs may play a role in the creation of beanbag chairs, or scrabble ledges, or paintings of people seated on chairs, or songs about letters seated on scrabble ledges, how do you objectively determine what is or is not a variant or descendent of any particular idea? Is the article I read by you a month ago an ancestor of this one? Is the fitness of the Mona Lisa the number of prints made of it, or the number of people who looked at it for at least one minute, or who like it, or who imitated it, or who at least once tried to smile like the Mona Lisa? We could decide that the fitness of an idea is a measure of not the number of 'offspring' it has, but its perceived value. However, that too is non-objective; the fitness of the idea for a new pesticide will have one value for its inventor, another for the farmer who uses it, another for the child who plays in trees sprayed with the pesticide, and so forth. Any measure of cultural fitness entails some degree of arbitrariness. As Vetsigian *et al.* (2006) put it, when ancestry arises through acquired change as opposed to inherited change, descent with variation is not genealogically traceable because change is not delimited to specific ancestors but affects the community as a whole.

## Natural Selection Requires Random Variation

The last misunderstanding addressed by Henrich et al is the belief that culture cannot be modeled as a process of natural selection because cultural variation is not generated randomly. They point to a statement by Pinker (1997) as a typical dismissal of the project: "a complex meme does not arise from the retention of copying errors... The value added with each iteration comes from focusing brainpower on improving the product, not from retelling or recopying it hundreds of thousands of times in the hope that some of the malaprops or typos will be useful." Henrich et al. (2008) argue that randomness is not necessary for the Darwinian framework to hold because "selective forces… require only variation not *random* variation" (p. 131).

The situation is actually subtler than either of these positions. It *is* possible for the theory of natural selection to be applicable even if the underlying process is not random. However, in that case, although not genuinely random, the process must be approximated by a random distribution. Systematic deviations from randomly generated variation render natural selection inapplicable as an explanatory framework because what is giving rise to change over time is the *nature* of those biases, *not* the mechanism Darwin identified—population-level change in the distribution of variants over generations of exposure to selective pressures. Biological variation is not genuinely random—for example, we can trace the source of some mutations to various causal agents—but the assumption of randomness generally holds well enough to serve as a useful approximation. Natural selection operates by generating *lots* of possibilities, through a process that can be approximated by a random distribution, such that by chance at least one of them is bound to be fitter than what came before.[8] Because of how human memory encodes information, in culture there is no need for large numbers of possibilities generated by chance processes. Culture works by generating *few* possibilities intelligently, such that they are more likely than chance to be adaptive. The ideas we generate reflect our knowledge, instincts, experiences, and how they are organized in our distributed, content-addressable memories, as well as the current context.

In sum, for natural selection to be applicable to the description of a given process there must be heritable variation with respect to innate (genetically mediated) traits that, if not random, can be approximated by a random distribution. *To the extent that variation is biased from a random distribution, what is giving rise to change over time is the nature of these biases, not natural selection*. With respect to culture, not only is there no mechanism for cultural inheritance, but the creative processes that fuel cultural change are goal-driven, intuitive, strategic, and forward-thinking, i.e. far from random.

## Toward a Theory of How Culture Evolves

Elements of culture certainly exhibit 'descent with modification', or incremental adaptation to the constraints of their environment, but difficulties have arisen in determining exactly in what sense they evolve. We began by addressing misunderstandings concerning *what* evolves through culture. We saw that mental representations are neither discrete nor continuous. Their structure reflects the distributed, content-addressable manner in which they are encoded in memory, and it is this that enables beliefs and ideas to change over time, and allows inferences and associations to be made, potentially giving rise to new ideas.

Examining the question of whether mental representations constitute replicators, we saw that Dawkin's notion of replicators captures some aspects of self-replication but ignores some of the most fundamental principles of natural selection. These include (1) a code that gets used both as self-assembly instructions and self-description, (2) a genotype/phenotype distinction, and (3) a lack of transmission of acquired characteristics, where (2) and (3) are a consequence of (1). The assumption that the replicator notion captures the essential elements of replication through natural selection has misleadingly led to support for the idea that culture evolves through natural selection.

Natural selection explains change in the frequency of inherited traits (those transmitted from one generation to another by way of a self-assembly code such as the genetic code), not acquired traits (those obtained between transmission events). Indeed natural selection is inapplicable whenever there *is* transmission of acquired characteristics. Attempts to force culture into a Darwinian framework (even those that pay lip service to the importance of a "rich psychology") leads to models in which "psychology" is reduced to sources of variation that occur during social transmission through processes such as copying error. Moreover, this variation must be randomly generated. If the above conditions—negligible transmission of acquired traits and random variation—are not met, the evolutionary process is explained by whatever is biasing or modifying the variation, not by natural selection, i.e. it is not due to statistical change in the frequency of heritable variations over generations due to differential response to selective pressure.

---

[8] Actually, some biological situations, such as assortative mating, cannot be accurately approximated by a random distribution, and to the extent that this is the case natural selection gives a distorted model.

Darwinian approaches to culture such as memetics and dual inheritance theory have made an important contribution in that by identifying where the analogy between biology and culture breaks down, we move forward toward a viable theory of how culture evolves. But Darwin's theory of natural selection is not appropriate for culture. Failure to appreciate Darwinian approaches to culture do not stem from a "tendency to think categorically rather than quantitatively" (p. 134) as Henrich et al. (2008) claim, nor are critics "not well equipped to digest mathematical models" (p. 121). Criticisms of these models reflect genuine limitations in the methods used and the theoretical foundation upon which they are based.

Research at the forefront of biology suggests that natural selection may be inapplicable to the early stages of *any* evolutionary process. Vetsigian et al. (2006) write: "the evolutionary dynamic that gave rise to translation [the process by which the proteins that make up a body are constructed through the decoding of DNA] is undoubtedly non-Darwinian" (p. 10696). Indeed in attempting to determine how culture evolves it is instructive to look at how early life itself evolved. Research into how life began is stymied by the improbability of a spontaneously generated structure that replicates using a self-assembly code (such as the genetic code). This has led to the widespread belief that the earliest self-replicating structures were autocatalytic sets of molecules that arose through self-organization, and evolved (albeit haphazardly) through horizontal (or lateral) transfer of innovation protocols, sometimes referred to as *communal exchange*[9] (Gabora, 2006; Kaufman, 1993; Vetsigian, Woese, & Goldenfeld, 2006; Weber, 1998, 2000; Weber & Depew, 1996; Williams & Frausto da Silva, 2003; Woese, 2002, 2004). In *communal exchange*, information is transmitted not by way of a self-assembly code from parent to offspring, but through encounters with others who may or may not be related. This communally exchanged information may be retained when the individual replicates. As Woese (2004) puts it: "Our experience with variation and selection in the modern context do not begin to prepare us for understanding what happened when cellular evolution was in its very early, rough-and-tumble phase(s) of spewing forth novelty" (p. 183). The work of Woese and his colleagues indicates that early life underwent a transition from a fundamentally cooperative process of horizontal evolution through communal exchange, to a fundamentally competitive process of vertical evolution through natural selection by way of the genetic code. This transition is referred to as the

---

[9] A biological example is horizontal gene transfer (HGT), or the exchange of simpler molecular elements by protocells (the earliest forms of life). A cultural example is giving directions to someone you meet on the street. The distinction between particulate and blended does not readily apply in communal exchange; ideas blend at the conceptual level, but at the level of microfeatures, but at the level of microfeatures they can behave in a discrete manner (see Gabora, 2010).

*Darwinian threshold* (Woese, 2002) or *Darwinian transition* (Vetsigian et al., 2006). Kalin Vetsigian (pers. comm.) estimates that the period between when life first arose and the time of the Darwinian threshold spanned several hundred million years.

If natural selection cannot explain how life arose, nor the earliest chapter of its evolutionary history, it is little wonder that efforts to apply it to culture have yielded little in the way of explanatory or predictive power. It seems sensible that if it took a few hundred million years for natural selection to emerge as the mechanism by which life evolves, it would not immediately be the mechanism by which culture evolves either. An alternative is that the evolution of culture is less akin to that of present-day, DNA-based life than to that of the earliest life forms (Gabora, 2000, 2004, 2008; Gabora & Aerts, 2009). The proposal is that what evolves through culture is *worldviews,* the integrated webs of ideas, beliefs, and so forth, that constitute our internal models of the world, and they evolve, as did early life, not through competition and survival of the *fittest* but through transformation of *all*. In other words, the assemblage of human worldviews changes over time not because some replicate at the expense of others, as in natural selection, but because of ongoing mutual interaction and modification. Elements of culture such as rituals, customs, and artifacts reflect the states of the worldviews that generate them. The theory is consistent with network-based approaches to modeling trade, artifact lineages, and the social exchange of knowledge and beliefs (e.g. Atran, Medin, & Ross, 2005; Lipo, 2006; Moody & White, 2003; White, Owen-Smith, Moody, & Powell, 2004). It is also more consistent than a 'survival of the fittest' perspective on culture with findings that human society is more cooperative than either expected utility or natural selection on genetic variation would predict (Henrich & Henrich, 2006; Richerson & Boyd, 1998; see also Boyer & Bergstrom, 2008). Let us move forward and see if this or some other theory can provide an explanatory framework by which we can scientifically understand the process by which we are, at an ever-increasing rate, transforming ourselves and our planet.

**Acknowledgements**
This work was funded in part by grants to the author from the *Social Sciences and Humanities Research Council of Canada* and the GOA program of the Free University of Brussels.

FIGURE CAPTIONS

Figure 1. This furniture inspired by the game of scrabble is an example of the ubiquitous phenomenon of blending of cultural 'lineages'.

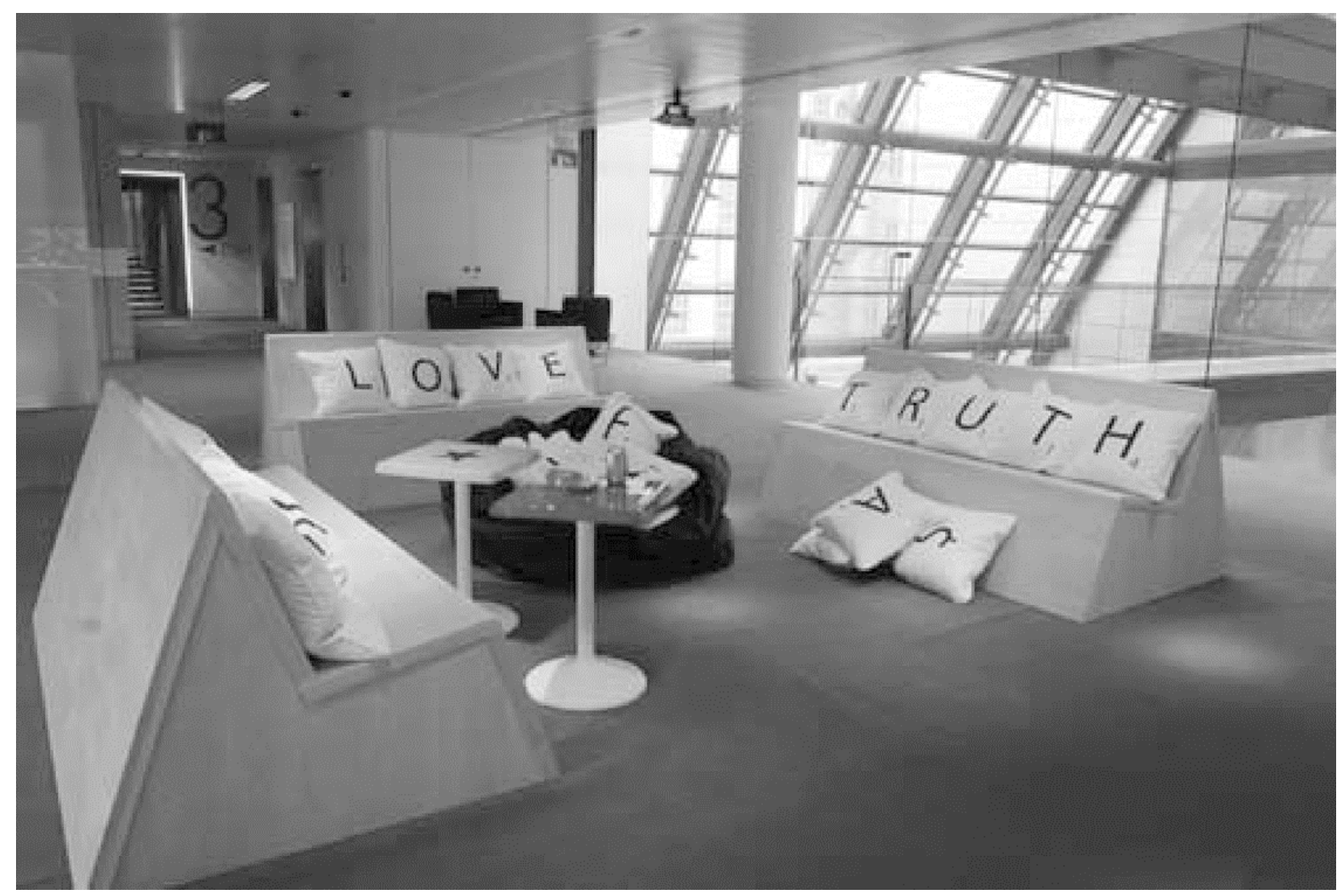

Figure 1.